\documentclass[aps,prb,twocolumn,superscriptaddress,eqsecnum]{revtex4-1}
\usepackage{amsmath}
\usepackage{color}
\usepackage{graphicx}
\usepackage{hyperref}
\usepackage{bm}
\usepackage{amsfonts}

\def\Re{{\rm Re}}
\def\Im{{\rm Im}}

\def\be{\begin{equation}}       \def\ee{\end{equation}}
\def\bea{\begin{eqnarray}}      \def\eea{\end{eqnarray}}
\def\ba{\begin{array} }
\def\ea{\end{array} }
\def\bnum{\begin{enumerate} }
\def\enum{\end{enumerate}}
\def\lf{\left}
\def\rt{ \right}

\def\=>{\Rightarrow}
\def\>{\rightarrow}

\def\eye2{Fathbb{I}}

\def\Q{\bm{Q}}

\def\d0{\Delta_{0}}

\newcommand{\beq}{\begin{equation}} 

\newcommand{\eeq}[1]{\label{#1} \end{equation}}

\newcommand\bmb{\left( \begin{matrix}}
\newcommand\emb{\end{matrix} \right)}
\renewcommand{\(}{\left(}
\renewcommand{\)}{\right)}

\renewcommand{\Im}{\mathrm{Im}}

\def\LSCO{La$_{2-x}$Sr$_x$CuO$_4$}
\def\LBCO{La$_{2-x}$Ba$_x$CuO$_4$}
\def\YBCO{YBa$_2$Cu$_3$O$_{6+x}$}

\def\BSCCO{Bi$_2$Sr$_2$CaCu$_2$O$_{8+\delta}$}

\def\C60{A$_x$C$_{60}$}

\def\HgCu3{HgCa$_2$Cu$_3$O$_{8+y}$}
\def\HgCu4{HgBa$_2$Ca$_3$Cu$_4$O$_{10+y}$}
\def\TlCu{Tl$_2$Ba$_2$CuO$_{6+\delta}$}
\def\TlCu3{Tl$_2$Ba$_2$Ca$_2$Cu$_3$O$_{10+y}$}
\def\TlCu4{Tl$_2$Ba$_2$Ca$_3$Cu$_4$O$_{12+y}$}

\def\BiCu3{Bi$_2$Sr$_2$Ca$_{2}$Cu$_3$O$_y$}

\def\8LSCO{La$_{1.88}$Sr$_{.12}$CuO$_4$}
\def\110LNSCO{La$_{1.5}$Nd$_{0.4}$Sr$_{0.1}$CuO$_{4}$}
\def\stage4LCO{La$_{2}$CuO$_{4+\delta}$}
\def\Y248{YBa$_2$Cu$_4$O$_8$}

\def\NbSe2{NbSe$_2$}
\def\TaSe2{TaSe$_2$}
\def\TiSe2{TiSe$_2$}
\def\NaCoOH2O{Na$_{0.3}$CoO$_{2y}$H$_2$O}
\def\MgB2{MgB${}_2$}

\def\URu2Si2{URu$_2$Si$_2$}

\def\Ba122{Ba(Fe$_{1-x}$Co$_x$)$_2$As$_2$}

\begin{document}
\title{Pair Density Waves in Superconducting Vortex Halos}
\author{Yuxuan Wang} 
\affiliation{Department of Physics and Institute for Condensed Matter Theory, University of Illinois at Urbana-Champaign, 1110 West Green Street, Urbana, IL  61801-3080, USA}
\author{Stephen D. Edkins} 
\affiliation{Department of Applied Physics, Stanford University, Stanford, CA 94305, USA}
\affiliation{LASSP, Department of Physics, Cornell University, Ithaca, NY 14853, USA}
\author{Mohammad H. Hamidian} 
\affiliation{Department of Physics, Harvard University, 17 Oxford St., Cambridge, MA 02138, USA}
\affiliation{LASSP, Department of Physics, Cornell University, Ithaca, NY 14853, USA}
\author{J.~C.~S\'eamus Davis} 
\affiliation{LASSP, Department of Physics, Cornell University, Ithaca, NY 14853, USA}
\affiliation{CMPMS Department, Brookhaven National Laboratory, Upton, NY 11973, USA}
\affiliation{School of Physics, University of St. Andrews, St. Andrews, Fife KY16 9SS, Scotland}
\author{Eduardo Fradkin}  
\affiliation{Department of Physics and Institute for Condensed Matter Theory, University of Illinois at Urbana-Champaign, 1110 West Green Street, Urbana, IL  61801-3080, USA}
\author{Steven A. Kivelson}
\affiliation{Department of Physics, Stanford University, Stanford, CA 94305, USA}

\date{\today }

\begin{abstract}
We analyze the interplay between a $d$-wave uniform superconducting 
  and a pair-density-wave (PDW) order parameter 
  in the  
  neighborhood of a 
  vortex. 
  We develop a phenomenological nonlinear sigma-model,  
  solve the saddle point equation for the order parameter configuration, and compute the resulting  local density of states in the vortex halo. 
  The intertwining of the 
  two superconducting orders leads to a charge density modulation with the same periodicity as  
the PDW, which is twice the period of the charge-density-wave that arises as a second-harmonic of the PDW itself.  We discuss key features of the charge density modulation 
  that can be directly compared with recent results from scanning tunneling microscopy 
  and speculate on the role PDW order may play in the global phase diagram of the hole-doped cuprates.
\end{abstract}
\maketitle

\section{Introduction}

In the conventional Bardeen-Cooper-Schrieffer (BCS) theory of superconductivity, electrons form Cooper pairs with zero total momentum, and the resulting superconducting (SC) order parameter is spatially uniform. In pair-density-wave (PDW) SC states, on the other hand, the momenta of Cooper pairs are nonzero, and the order parameter field $\Delta(\bm r)$ is nonuniform and oscillatory in space.
Spatially non-uniform superconducting states were first considered long ago by Fulde and Ferrell \cite{Fulde-1964} (FF) and Larkin and Ovchinnikov \cite{Larkin-1964} (LO) in a BCS model with a uniform external magnetic
 (Zeeman)
  field.   More recently, PDW 
  states have been proposed\cite{Berg-2007,berg-2009,agterberg-2008,fradkin-2014,lee-2014,Pepin-2014,Wang-2015,Wang2-2015,Loder-2010} to exist \emph{in the absence} of an external magnetic  field in a family of cuprate high-temperature superconductors (HTSC).  In several numerical studies PDW states have been shown to be close competitors to the ground state of 2D $t-J$ models with a uniform d-wave superconductor (over a broad range of parameters).\cite{Himeda-2002,Raczkowski-2007,corboz-2014}

Specifically, 
a  dynamical layer decoupling is observable in the  transport properties\cite{Li-2007,Tranquada-2008,Rajasekaran-2018}   
of {\LBCO}  
with doping concentrations less than or equal to $x=1/8$, both 
in the presence and absence of a magnetic field, and in underdoped {\LSCO}~\cite{abbamonte-2005,lake-2002,Schafgans-2010,Schafgans-2010b,Tranquada-2015}  {and La$_{2-x}$Ca$_{1+x}$Cu$_2$O$_6$~\cite{Tranquada-2018}}
in large enough magnetic fields.
This can be naturally explained by a PDW SC state, and thus constitutes dramatic, albeit indirect evidence of the existence of a PDW in the ``214 family'' of HTSCs.
 Whether PDW-type phases arise in HTSC other than the
 {lanthanum-based 214} materials is presently not known, although certain indirect evidence of its existence in both {\BSCCO}\cite{Hamidian-2016,lee-2014} and {\YBCO}\cite{ong-2016,lee-2014} has been adduced by several authors.
Evidence for PDW-type SC states has also been found in the heavy-fermion material CeRhIn$_5$~\cite{Park-2012} {and CeCoIn$_5$~\cite{Kenzelmann-2013}}  at high magnetic fields.

Charge-density-wave (CDW) correlations, similar to 
those associated with the well-known\cite{kivelson-2003} ``stripe'' order  
have  
recently been discovered in  all families of hole-doped HTSCs in which the requisite experiments have been carried out, including
 in {\YBCO},\cite{Wu-2011,chang-2012,Ghiringhelli-2012,Wu-2013,blackburn-2013,Blanco-Canosa-2014,Comin-2015,Achkar-2016,Jang-2016} in {\BSCCO},\cite{lawler-2010,silva_neto-2012, Fujita-2014,Silva_Neto-2014,Hamidian-2015,Vig-2015,Mesaros-2016} and in HgBa$_2$CuO$_{4+\delta}$.\cite{Greven-2014}
This order is characterized by an order parameter, ${\rho}_{\bm K}$, where ${\bm K}$ signifies the CDW ordering vector that appears to vary somewhat with doping concentration and with the particulars of the crystal structure, but which typically corresponds to a period 
$\sim \ 3a_0$ - $4a_0$,\cite{Mesaros-2016} where $a_0$ is the lattice constant in $xy$ directions.  The CDW order is 
 short-range correlated in zero field, but tends to strengthen and develop substantially longer correlations in a high magnetic field.  

Despite the fact that the CDW order has a reasonably high onset temperature, and that it apparently competes on an equal footing with the SC order, in many ways the CDW order appears to be extremely weak -- which is a partial explanation for the fact that it was so difficult to detect for so many years.  This has lead to the suggestion that it might be a parasitic order, associated with a different, possibly stronger ordering tendency.
One important feature of the PDW state is that 
it generally induces an associated (composite) CDW order
$
\tilde{\rho}_{2\bm Q} = B_2\Delta_{-\bm Q}^*\Delta_{\bm Q},
$
where $\Delta_{\bm Q}$ is the PDW order parameter, ${\bm Q}$ is the PDW ordering vector, and $B_2$ is an appropriate coupling constant.  More generally, in the presence of both CDW and a PDW order, there is a natural tendency for a mutual commensurate lock-in (encoded in a specific cubic term in the Landau theory\cite{zachar-1998}) which favors the condition  ${\bm K}=2{\bm Q}$.  
On the other hand, when PDW and uniform SC coexist, the intertwining of the two orders leads
 {\it inevitably} to an additional charge density modulation $\tilde{\rho}_{\bm Q}$ with the {\em same} wave vector as that of the PDW given by
$\tilde{\rho}_{\bm Q} = B_1\Delta_0^*\Delta_{\bm Q}$,
where $\Delta_0$ is the uniform SC order parameter.  In other words, if   the ${\bm K}=2{\bm Q}$ CDW  has a $4a_0$ periodicity, then
the period of  
the charge modulation where uniform and PDW order coexist would be  $8a_0$.
While it is difficult to directly detect PDW 
order,~\cite{pdw-higgs} the observation of such a  ``$8a_0$" charge modulation would provide strong evidence for the PDW state in HTSC materials.

Since the 8$a_0$ charge modulation is a composite of SC and PDW order parameters, its magnitude is strongest when the SC and PDW are of comparable strength. However, in the ground state of a HTSC without any defects, one expects that the competition between SC and PDW strongly reduces the latter or 
completely eliminates it. One way to overcome this issue is to look for signatures of the $8a_0$ charge modulation near a SC vortex 
core, where the SC order parameter {is suppressed.~\cite{agterberg-2008,Agterberg-Garaud-2015}}
 In this paper we focus on the structure of the PDW order parameter and its accompanying charge modulation in a vortex core halo. We show that the PDW order and the  resulting $8a_0$ charge modulation is enhanced near a SC vortex. 
The results we obtain here can be directly compared with the measurement of local density of states (LDOS)  in STM.  

In this paper we develop a phenomenological nonlinear sigma-model to describe the intertwining of the PDW and SC order parameters, both with $d$-wave form factors.
 We analyze various coupling terms in this effective model, and solve the saddle point equation in a vortex configuration for the SC order parameter. By an energetic argument, we show that the PDW order parameter  with a \emph{uniform} phase is enhanced near the vortex core, even in the presence of a phase coupling term that by itself favors a phase winding configuration of the PDW order. 
To facilitate comparisons with STM experiments, we compute the local density of states (LDOS) using parameters relevant for a typical cuprate HTSC material. 
 The induced CDW order, both at ${\bm Q}$ and $2{\bm Q}$, have dominantly $s$-wave form factors.
 Moreover,  
 since the phase of the SC order parameter changes by $\pi$ across the vortex center, 
 the ${\bf Q}$ charge modulation patterns across the vortex center 
 must exhibit a $\pi$ phase shift.

The present study was motivated in part by a recent  (unpublished) STM study\cite{STM-8a} of slightly underdoped
 {\BSCCO} in which a period $8a_0$ CDW has been observed in the halo of  field-induced vortices.  Broadly, the measured characteristics of the field-induced CDW are consistent with the theoretical results we have obtained.  {\em This constitutes dramatic affirmative evidence that a PDW phase is part of the physics of the cuprate HTSCs}.
 
The remainder of this paper is organized as follows: In Sec.\ \ref{sec:nlsm} we develop an effective theory for a cuprate HTSC system with intertwined PDW and SC orders. We show that this model naturally incorporates different types of charge density modulations as parasitic order parameters, and we discuss its interplay with an independent 
(predominantly d-wave form factor) CDW degree of freedom.  In Sec.\ \ref{sec:saddle} we solve the saddle point equation given by the effective theory for the configuration of the order parameters with the boundary condition enforced by an isolated SC vortex. 
In Sec.\ \ref{sec:ldos}, we couple this order parameter configuration to a cuprate-like Fermi surface and compute the 
LDOS. 
 In Sec.\ \ref{sec:phasediagram} we 
speculate more broadly about the significance of the observation of PDW order in vortex cores for the physics of the cuprates.  Here, we propose a plausible phase diagram as a function of temperature and magnetic field and relate it to a variety of other experiments that have shed light on the phases that arise as superconductivity is suppressed.  Here we also discuss the similarities and differences of the present results with the earlier ground-breaking work of  Ref.~\onlinecite{Agterberg-Garaud-2015}, in which a similar study of PDW in vortex cores was carried out, but with importantly different underlying physical assumptions and broader consequences.
In Sec.\ \ref{sec:discussion} we  summarize our key findings, and in the Appendix, we discuss a variety of different possible microscopic subtleties of the vortex core structure, including consideration of the case in which the PDW order has a vortex centered at the same position as the uniform SC.

\section{Effective theory for PDW and SC}
\label{sec:nlsm}
The Landau-Ginzberg effective field theory for intertwined uniform SC and PDW orders is shown to quartic order in Ref. [\onlinecite{fradkin-2014}].  This approach makes the symmetries apparent, but the expansion in powers of the order parameter is only valid in the vicinity of a high order multi-critical point.  In contrast, we will be primarily interested in the behavior of the system at low $T$,  mostly deep inside the uniform SC phase.

At a microscopic level, it is reasonable to think that since the PDW and the uniform SC order involve precisely the same electrons 
they compete
ferociously. 
 Indeed, at short distances, (which is the relevant scale near an isolated vortex), $d$-wave SC order and PDW order with $d$-form factor should behave very similarly. 
Thus, it is suitable to start by considering a nonlinear sigma-model that enforces a fixed magnitude of the local pair-field, without distinguishing between the uniform SC and 
PDW components.
The resulting theory has an unphysical large (SO(10)!) symmetry, but this can be corrected by including appropriate explicit symmetry breaking terms, under the assumption that these are in some sense small.  For a tetragonal system, the exact symmetries that remain are $U(1)\times U(1)\times U(1)\times C_4$, where the first $U(1)$ is associated with charge conservation, there is a $U(1)$ symmetry corresponding to translational symmetry in the $x$ and $y$ directions, and the $C_4$ symmetry reflects the assumed point group-symmetry.
  In this model, charge density modulations do not appear as separate degrees of freedom, but as we shall see, they emerge naturally as 
  composite orders.  

We define a five-component complex order parameter field
\be
\Phi =\langle \Delta,\Delta_{\Q},\Delta_{-\Q}, \Delta_{\Q^\prime},\Delta_{-\Q^\prime}\rangle
\ee
where $\Delta$ is the uniform $d$-wave SC order parameter, $\Delta_{\Q}$ is the PDW order parameter with $d$-wave form factor and an 
ordering wave-vector $\Q$, $\Q^\prime$ is related to $\Q$ by a $C_4$ rotation, and there is a constraint that 
\be
|\Phi|^2=1.
\ee
The model we consider consists of the reference SO(10) non-linear sigma-model plus symmetry breaking terms. {In this work we will be interested in the case in which the order parameters are static and thus will be treated as classically spatially-varying fields. The free energy of a general  configuration of the order parameter fields is}
\be
\label{nlsm}
S_{\rm I}[\Phi] = \int d^2x \left [\ \frac \kappa 2 |\vec D \Phi|^2  + \delta {\cal L}\  \right]
\ee
where $\kappa >0$, $\vec D = \vec \nabla - 2e i\vec A$,
and
\bea
\label{dotdotdot}
\delta {\cal L} =&& -\frac \epsilon 2 |\Delta|^2 {+\frac {
\delta\kappa} 2 |\vec D \Delta|^2 } \nonumber \\
&&+ \frac {\tilde \kappa}2\left[ |\hat \Q \cdot \vec D \Delta_{\Q}|^2 + |\hat \Q \cdot \vec D \Delta_{-\Q}|^2 \right . \\
&&\left . + |\hat \Q^\prime \cdot \vec D \Delta_{\Q^\prime}|^2 + |\hat \Q^\prime \cdot \vec D \Delta_{-\Q^\prime}|^2\right] \nonumber \\
&&+\gamma\left[( |\Delta_{\Q}|^2-|\Delta_{-\Q}|^2)^2 +  (|\Delta_{\Q^\prime}|^2-|\Delta_{-\Q^\prime}|^2)^2\right] \nonumber \\ 
&& + \tilde \gamma \left(|\Delta_{\Q}|^2+|\Delta_{-\Q}|^2 \right)\left(  |\Delta_{\Q^\prime}|^2+|\Delta_{-\Q^\prime}|^2\right)\nonumber \\
&&-\lambda \left[ e^{i\delta}\Delta^*\Delta^*( \Delta_{\Q}\Delta_{-\Q} + \Delta_{\Q^\prime}\Delta_{-\Q^\prime})\right] + c.c. \nonumber \\
&&- \tilde \lambda \left[e^{i\tilde\delta} \Delta^*_{\Q}\Delta^*_{-\Q} \Delta_{\Q^\prime}\Delta_{-\Q^\prime}\right] + c.c. + \ldots
\nonumber
\eea
where we have exhibited explicitly all the necessary terms for present purposes, while $\ldots$ represents additional terms that are less important and will not be treated explicitly.  The terms proportional to $\tilde \kappa$ 
(which can be of either sign so long as {$\kappa + \tilde \kappa >0$)}
link the exchange of components $\Delta_{\Q}$ and $\Delta_{\Q^\prime}$ with the interchange of the directions of the associated ordering vectors.  We will assume that the remaining couplings (which in general can be of either sign) are all positive.  Thus, the term proportional to $\epsilon$ is the leading term that favors uniform SC order over PDW;   the term proportional to $\delta\kappa$ (subject to the condition $\kappa +\delta \kappa>0$) is an example of a  potentially important additional term that reflects the difference between the two types of order parameter;
 the term proportional to $\gamma$
favors the LO over  FF states ( i.e.,  favors the state in which  $|\Delta_{\Q}|=|\Delta_{-\Q}|$; the term proportional to $\tilde \gamma$ favors  PDW stripes over checkerboards;  the terms proportional $\lambda$ and $\tilde \lambda$ couple the relative superconducting phases of the various SC orders to a value determined by $\delta$ and $\tilde \delta$.~\cite{agterberg-2008} The terms describing the coupling of the order parameters to the fermions are presented and discussed in Sec.\ \ref{sec:ldos}.

Since the PDW order breaks translational symmetry, it induces charge density modulations as composite orders. In particular, 
to quadratic order there are 
charge modulations with ordering momenta $\bm Q$ and $2 \bm Q$, 
\be
{\tilde \rho_{\Q}(\bm r) = B_1 \(\Delta^*\Delta_{\Q} + \Delta\Delta^*_{-\Q}\)/2}
\label{eq:5}
\ee
and
\be
\tilde \rho_{2\Q}(\bm r) =B_2 \Delta^*_{-\Q}\Delta_{\Q}.
\label{eq:6}
\ee

As we are considering an extreme type-II superconductor, we set $\vec A=\vec 0$;  we will in the next section focus on a single vortex by imposing boundary conditions at large distances $\Phi(\bm r) \sim e^{i\theta}\langle 1,0,0,0,0\rangle$.

\subsection{Coupling to an independent CDW order}

From a macroscopic perspective, the ``minimal model" above includes all the relevant phases, in which the charge density modulations are given 
composite order parameters. 
From this perspective, the CDW order  observed in the pseudogap region might be considered as a signature of ``vestigial order''\cite{nie-2013} which persists above the temperature at which the expectation values of the primary order parameters vanish, and the CDW ordering- vector ${\bm K}$ is interpreted as  
${\bm K}=2\bm Q$.

However, 
 as we will discuss at greater length in Sec. \ref{sec:phasediagram}
  for a general description 
  of the cuprate phenomenology, it is  useful to include
  an independent CDW order parameter with wave-vector $\bm K$. 
  In the first place, if the CDW were purely parasitic on the PDW, one would expect that 
superconducting correlations 
would necessarily also be substantial in the entire regime 
  in which 
  CDW correlations are observed. Although there is currently no convincing theory of such fluctuational regimes, it is natural to presume that such PDW fluctuations would lead to dramatic and detectable consequences, including a large contribution to a SC-like fluctuation conductivity. There are clear and strong SC fluctuation effects seen within maybe 30K above $T_c$, but 
  for temperatures further above $T_c$ (in the regime where substantial CDW correlations are still detected) such effects are exceedingly weak.  Second, it has been established~\cite{Fujita-2014,Sachdev-2013,Wang-2014} that the $2\bm Q$ CDW has a predominantly $d$-wave form factor. However, a parasitic CDW order parameter $\tilde\rho_{2\bm Q}$ should have a predominantly $s$-wave form factor, simply because it goes as the square of the PDW order. It is important to point out here that the $s$-wave and $d$-wave form factors are not true symmetry designations, since all symmetries that can be used to distinguish the two are broken by the 
  ordering wave-vector $\bm K$. However, to the extent that this distinction remains \emph{approximately} valid, it still argues for introducing a separate CDW order parameter  $\rho_{2\bm Q}$ with predominantly $d$-form factor, which is weakly coupled to $\tilde\rho_{2\bm Q}$.

Following this argument, we introduce the effective model that incorporates independent CDW degrees of freedom as
\bea
S_{\rm II}[\Phi,\rho] = &&S_{\rm I}[\Phi] + S_{cdw}[\rho] +S_{dis}\\
\label{lockin}
&&+ \alpha\int  d^2x\left[\rho_{2\Q}^* \tilde \rho_{2\Q} +\rho_{2\Q^\prime}^* \tilde \rho_{2\Q^\prime} + c.c.\right]
\nonumber
\eea
where we take the coupling $\alpha$ to be small (owing to the $d$-form factor of $\rho_{\Q}$), $S_{cdw}$ is the pure CDW part of the {free energy}, and $S_{dis}$ incorporates all spatially non-uniform contributions to the effective free energy from ``disorder.''   The leading order coupling to $\rho$ is a ``random field'' coupling to quenched disorder.  There are higher-order disorder couplings to the superconducting fields and, since in some cases impurities are seen in the cuprates to ``punch a hole in the superconductivity,'' we consider terms of the form
\be
S_{dis} =  \int d^2x \left\{ [v \rho^*_{\Q} + v^\prime \rho_{\Q^\prime} +  c.c.] +u|\Delta|^2\right\} + \ldots
\ee
where $v(\vec x)$ and  $v^\prime(\vec x)$ are complex random potentials and $u(\vec x )$ is a real random potential.

The effective free energy $S_{\rm II}$ is relevant for analyzing the impurity effects on PDW and charge modulations near impurities. Indeed, near an isolated impurity, $v$ and $v^\prime$ are non-zero over some finite range, which tends to induce CDW order, even if it is somewhat suppressed in the disorder-free  $S_{cdw}$.   Some impurities also suppress uniform SC, i.e., have a large positive value of $u$, as suggested by STM data.\cite{STMnearZn,STMnearZn2}  It would be interesting to compare the results from $S_{\rm II}$ with STM measurements in the presence of isolated impurities. However, 
to study the properties an isolated superconducting vortex at temperatures well below the superconducting $T_c$ 
 we assume that the independent CDW order parameter  
is negligible and proceed to analyze  $S_{\rm I}[\Phi]$.

\section{saddle point equation near a SC vortex}
\label{sec:saddle}

For simplicity, we focus on the case in which 
{$\delta \kappa=0$} and $\tilde \kappa=0$, and look for a single vortex solution.  We thus {consider configurations that satisfy the} boundary conditions,
\be
\Phi(\vec x) \to e^{i\theta(\vec x)} \langle 1,0,0,0,0\rangle \ \ {\rm as } \ \ |\vec x| \to \infty
\ee
and search for the minimum energy solution,
\be
\frac {\delta S_{\rm I}} {\delta \Phi} = 0.
\ee

Specifically, we consider a case in which $\gamma$ and $\tilde \gamma$ are assumed to be large and positive. 
 In this case, the term in Eq.\eqref{dotdotdot} proportional to $\tilde \lambda$ is zero. {Consequently, we} look for solutions of the form
\bea
&& \Delta(\vec x)= e^{i\phi_{sc}(\vec x)} F(\vec x) \\
&&\Delta_{\Q}(\vec x)= e^{i[\phi_{cdw}
+\phi_{pdw}(\vec x)
]/2} G(\vec x)/\sqrt{2} \nonumber \\
&&\Delta_{-\Q}(\vec x)= e^{i[-\phi_{cdw}
+\phi_{pdw}(\vec x)
]/2} G(\vec x) /\sqrt{2} \nonumber \\
&&\Delta_{\Q^\prime}(\vec x)=  \Delta_{-\Q^\prime}(\vec x)=0\nonumber \\
&& G(\vec x) = \sqrt{1-F^2(\vec x) }\ \ {\rm where} \ \ 1 \geq F(\vec x) \geq 0\ . \nonumber
\eea 
(Of course, we could have chosen a PDW in the $\Q^\prime$ direction just as well.)  

There are three phase degrees of freedom of any such solution:  One of these correspond to an exact symmetry {(the global gauge symmetry), but its behavior at infinity is fixed by our assumed boundary condition} to be $\phi_{sc}(\vec x) = \theta(\vec x)$.   A second corresponds to translation of the charge density modulation, $\phi_{cdw} \to \phi_{cdw}+ \Q \cdot \vec \ell$,  which is not truly a symmetry, since the vortex core breaks translational symmetry. However, it is unclear how exactly the charge density modulation couples to the potential introduced by a vortex core,
 and for now we 
assume the pinning effect to be negligible and treat translation
  as an exact symmetry. Thus, at this level of analysis,  $\phi_{cdw}$ is arbitrary. The third involves the   phase of the PDW, 
   {$\phi_{pdw}(\vec x)$,} 
   relative to the phase of the uniform SC order  - this symmetry is broken by the term proportional to $\lambda$, and we will focus on the effect of this term later.

We first look for solutions {to the field equations that minimize $S_{\rm I}$ under the assumption that  the PDW order has a uniform phase 
 {\it i.e.} we take $\phi_{pdw}(\vec x)$ to be a position independent constant, $\phi_{pdw}$.  We show
will  that in this 
case,  the PDW component is peaked at the vortex core, similar to the meron solution~\cite{Gross-1978,Affleck-1986} for an $O(3)$ nonlinear sigma-model.  At the present level of approximation, $\phi_{pdw}$ and $\phi_{cdw}$ are arbitrary;  additional terms would be needed in the effective free energy to determine their value.  For simplicity, in the present section we set $\phi_{pdw}=0$ and choose $\phi_{cdw}$ so that $\Delta_Q(\vec x)$ is real.}

For this configuration, the expectation value for the term proportional to $\lambda$ in Eq.\eqref{dotdotdot} is identically zero because of the phase winding of the SC order parameter. We define the field $\vec n$,
\be
\vec n=(\Re\Delta (r), \Im \Delta (r), \sqrt{2}\Delta_Q).
\ee
It is straightforward to show that, for this class of solutions, the effective free energy is the same as that of an O(3) nonlinear sigma-model with Ising anisotropy
given by
\be
S_{\rm I}[\vec n] = \int d^2x \lf[\frac{\kappa}{2}|\nabla \vec n|^2 + \frac{\epsilon}{2} n_3^2 + \ldots\rt].
\ee
The anisotropic term favors a meron solution,\cite{Gross-1978,Affleck-1986} for which $\vec n$ lies in-plane at $|\vec x|\to \infty$ and points along $\hat z$ direction at $\vec x=0$. We make the following parametrization 
\be
\vec n = (\sin \alpha \cos \beta, \sin\alpha \sin\beta, \cos\alpha),
\ee
where $\cos\alpha(\vec x)=G(\vec x)$, and $\sin\alpha(\vec x)=F(\vec x)$.
 Since we have neglected the $\tilde \kappa$ terms, the solution is rotational invariant, and $\alpha= \alpha(|\vec x|)$ and $\beta=\theta(\vec x)$.
 The effective free energy becomes~\cite{Affleck-1986}
\begin{align}
S_{\rm I}[\alpha,\beta] =& \pi\kappa \int r dr \lf[ \lf(\frac{d\alpha}{dr}\rt)^2 + \frac{\sin^2 \alpha}{r^2} +\frac{\epsilon}{\kappa} \cos^2\alpha \rt]\nonumber\\
 =&\pi\kappa \int dt \lf[ \lf(\frac{d\alpha}{dt}\rt)^2 + {\sin^2 \alpha} + 
  e^{2t} \cos^2\alpha \rt],
\label{eq16}
\end{align}
where in the last step we defined $t=\ln (r/r_0) \in (-\infty, \infty)$ and 
the scale is defined as 
\be
r_0
\equiv \sqrt
{{\kappa}/{\epsilon}}.
\label{r0}
\ee
This free energy is minimized by solutions of
\be
\label{eq17}
\ddot \alpha  = 
{\frac{1}{2}}\left[1 - 
e^{2t}\right]\sin 2\alpha
\ee
 subject to the boundary condition $\alpha( t) \to 0$ as $t \to -\infty$ and $\alpha(t) \to \pi/2$ as $t \to +\infty$.  Manifestly, $r_0$ determines the size of the vortex halo.
Note that, 
without the $\epsilon$ term, the effective free energy is independent of the scale, $r_0$, and the free energy would be minimized by   a skyrmion configuration without a typical scale---i.e. the effective free energy is scale invariant.   

 We numerically solved Eq.\ \eqref{eq17} by mapping the effective free energy to the action of a classical point particle in a time-dependent potential, and we plot the magnitudes of PDW, SC and the resulting charge density modulation in Fig.\ \ref{fig:1}. Indeed, we see that the size of the vortex halo is given by Eq.\ \eqref{r0}. From Eqs.\eqref{eq:5} and \eqref{eq:6}, we compute the magnitudes of $\tilde\rho_{\bm Q}$ and $\tilde\rho_{2\Q}$ with $B_1=B_2=1$, which are also shown in Fig.\ \ref{fig:1}. Note that since $\tilde\rho_{\bm Q}$ is linear in the PDW order parameter $\Delta_{\Q}$ and $\tilde\rho_{2\bm Q}$ is quadratic, the latter decays faster in space away from the vortex center. In other words, $\tilde\rho_{\bm Q}$ has a larger correlation length than $\tilde\rho_{2\bm Q}$.

 The energy of such a meron is infrared divergent, given by
 \be
 \label{eq21}
 E=\pi\kappa\ln\frac{R_0}{r_0} = E_{vor} + \Delta E
 \ee
 where $R_0$ is the system size,  {and the creation energy of   a bare vortex (without an induced PDW in the core) is
 $E_{vor} =\pi \kappa \ln({R_0}/{a})$,
  where $a$ is an ultra-violet cutoff of order the superconducting coherence length;  the vortex halo in effect reduces the vortex core energy by 
  \be
  \Delta E = - \pi\kappa \ln\frac{r_0}{a}.
  \label{eq19}
  \ee
  
So far we have focused on a configuration where PDW order 
 has a uniform phase at the vortex core.
 {Since the term proportional to $\lambda$ in Eq.\ \eqref{dotdotdot} is minimized
 when  the phases of $\Delta_{\pm Q}$ are locked to that of $\Delta$,  it is also interesting to search for solutions of the saddle point equation with} phase locking between $\Delta$ and $\Delta_{\pm Q}$, {{\it i.e.} when there are coincident vortices in both the uniform SC and the PDW components of the order.} {We have also computed the order parameter structure and free energy of such a configuration. We found that the free energy of this configuration is (logarithmically) singular in the ultraviolet cutoff $a$, and hence is only a metastable solution. We show the details of this phase-locked solution in Appendix \ref{app:lock}.}

 \begin{figure}[hbt]
 \includegraphics[width=\columnwidth]{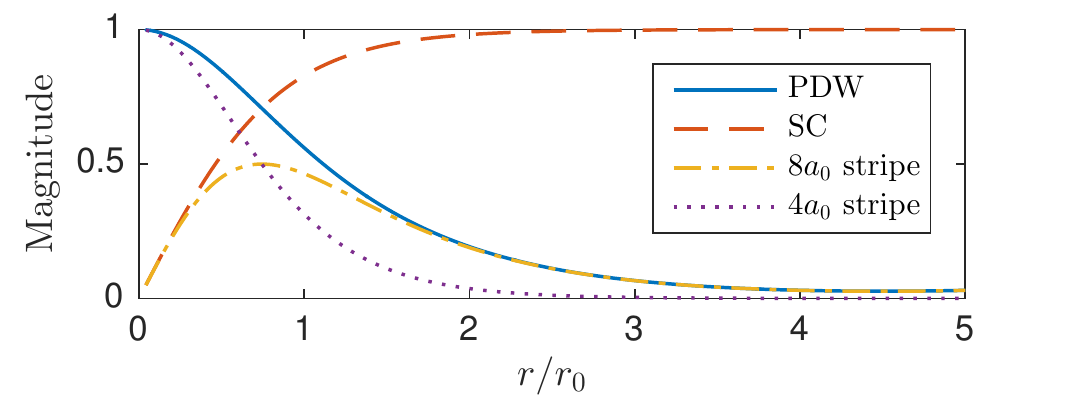}
 \caption{The numerical solution to the meron-like configuration of the order parameters 
{ in the vicinity of an isolated vortex.} {The PDW order parameter has a uniform phase while the SC one has a $2\pi$ winding.} The magnitudes of $\tilde\rho_{\bm Q}$ and $\tilde\rho_{2\bm Q}$ ({\it i.e.}, the period $8a_0$ and period $4a_0$ stripes) are computed from Eqs.\eqref{eq:5} and \eqref{eq:6} with coefficients $B_{1}=B_2=1$.}
  \label{fig:1}
 \end{figure}

\section{Local density of states 
in a vortex halo}
\label{sec:ldos}

\begin{figure}[hbt]
 \includegraphics[width=\columnwidth]{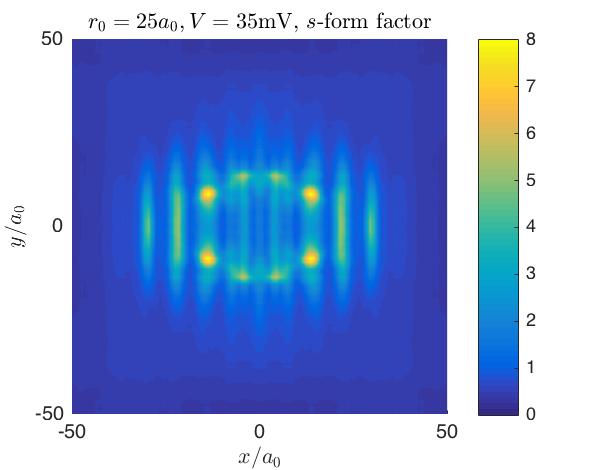}
 \caption{LDOS with $s$-form factor as a function of position at $V=35$mV with a vortex centered at the origin and a halo 
 size $\sim 25a_0$. {Note that even if the PDW and SC order parameters we used are rotationally invariant, the envelop of the charge density modulations are elongated along $x$-direction. }
}
  \label{fig:2}
 \end{figure}
 
 \begin{figure}[hbt]
 \includegraphics[width=\columnwidth]{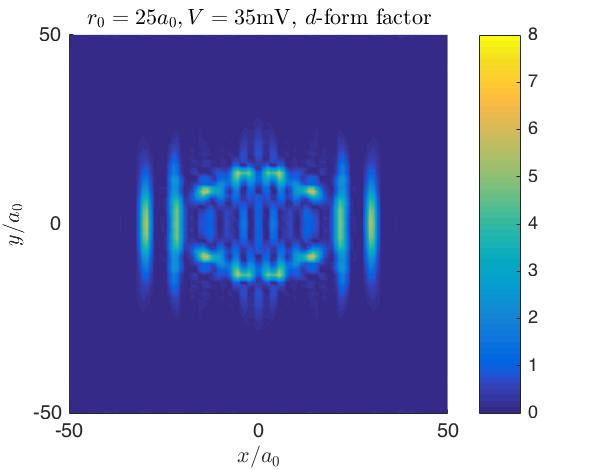}
 \caption{Absolute value of LDOS with $d$-form factor at $V=35$mV as a function of position with a vortex centered at the origin and a halo size $\sim 25a_0$. The color code scale is the same with that of Fig.\ \ref{fig:2}.
 }
  \label{fig:3}
 \end{figure}

We now address the implications of the above analysis for microscopic and spectroscopic properties of the system.  Specifically, we use the field configurations derived from the effective field theoretic considerations above as the input to  an appropriate Bogoliubov-de Gennes (BdG) effective Hamiltonian for the electrons.  
  Consider the 
  effective {lattice} Hamiltonian:
\bea
H_{tr}= && - \sum_{\bm r,\bm r^\prime,\sigma} t(\bm r-\bm r^\prime) c_{\bm r,\sigma}^\dagger c_{\bm r^\prime,\sigma}  \\
&&+\sum_{\bm r,\bm r^\prime}\left[ \Delta(\bm r,\bm r^\prime) c_{\bm r,\uparrow}^\dagger c_{\bm r^\prime,\downarrow}^\dagger + h.c.\right]
\nonumber
\eea
where we will take  
as an ansatz
\bea
\Delta(\bm r,\bm r^\prime)= &&\Delta_0D(\bm r-\bm r^\prime)\left [F(\bm r) e^{i\theta(\bm r)}\right . \\
&&\left . + G(\bm r) \cos(\Q \cdot \bm r + \phi_{{cdw}}){e^{i\phi_{pdw}}} \right]
\nonumber
\eea
where $D(\bm r)$ is the nearest-neighbor $d$-wave form factor and 
 the real functions, $F$ and $G$ are subject to the boundary conditions $F(\vec 0) =0$ and $F(\bm r) \to 1$ as $|\bm r|\to \infty$ and $G(\bm r) \to 0$ as $|\bm r|\to \infty$.

Given $H_{tr}$, we compute the local density of states near an isolated vortex, as this can %
 be directly compared to what is seen in an STM experiment.  
 We also compute the induced charge density modulation -- which could in principle be seen 
 with REXS.

 \begin{figure}[hbt]
 \includegraphics[width=\columnwidth]{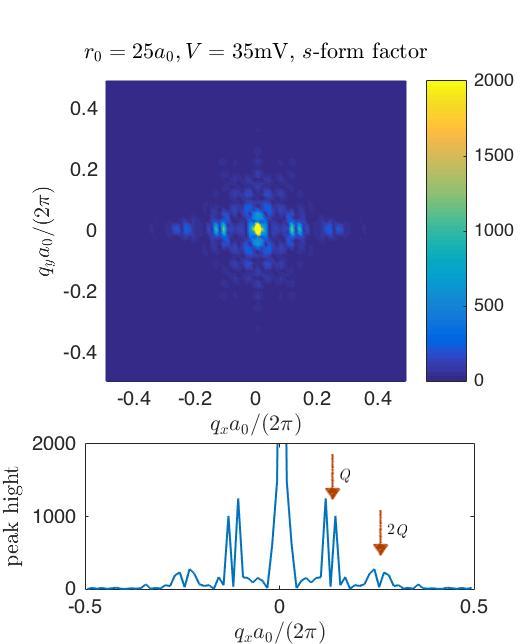}
 \caption{Top panel: Fourier transform of LDOS with $s$-form factor at $V=35$mV for a vortex halo of size $\sim 25a_0$. The peak at $Q=\pi/4$ is visible. Bottom panel: the same data along the $q_y=0$ cut.}
  \label{fig:4}
 \end{figure}
 
 \begin{figure}[hbt]
 \includegraphics[width=\columnwidth]{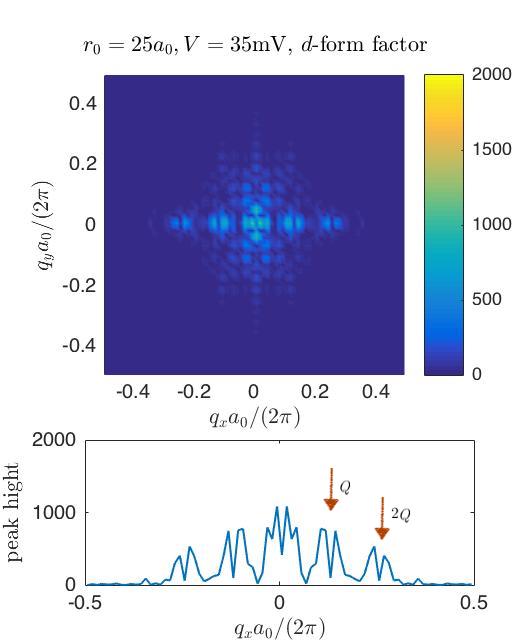}
 \caption{Top panel: the Fourier transform of LDOS with $d$-form factor at $V=35$mV for a vortex halo of size $\sim 25a_0$. The color code scale is the same with that of Fig.\ \ref{fig:4}. Bottom panel: the same data along $q_y=0$ cut.}
  \label{fig:5}
 \end{figure}
 
We have performed a direct diagonalization of the real space BdG Hamiltonian on a {$120a_0\times 120a_0$} lattice, and 
computed the LDOS 
with an energy resolution at $1$meV.  The period of PDW order parameter is set at $8a_0$.
For the normal state band structure, we have used a parametrization~\cite{zx11} of the band structure of {\BSCCO} in a single-band $t-t'-t''$ model with $t=0.22$eV, $t'=-0.034$eV,  $t''=0.036$eV, and $\mu=-0.243$eV. 
We 
take $\Delta_{0}
=40$meV, and the radius of the vortex halo (meron) 
to be $100$A, which is about $25a_0$. {Specifically, we will use the optimal solution (shown in Fig.\ \ref{fig:1}) from previous section as input for $F$ and $G$. In that solution, the phases $\phi_{cdw}$ and $\phi_{pdw}$ can be arbitrary. For definiteness, we take $\phi_{cdw}=\pi/2$, and $\phi_{pdw}=0$ for the following analysis. We present the results for other choices of $\phi_{cdw}$ and $\phi_{pdw}$ in Appendix \ref{app:cosQr}, and show that all the qualitative features are the same.}

We placed a single vortex on a torus, which means there are branch cuts of the order parameters far away from the vortex core at the unphysical edges of the torus.  We found these ``rough edges" generally do not affect the LDOS near the vortex much. When performing a Fourier transform, we exclude these rough edges.

In Fig.\ \ref{fig:2}, we show such a LDOS plot with $s$-form factor (probing on-site density) at $V=35$mV, at which the signal appears to be strongest.  Charge density modulations with period $8a_0$ are clearly discernible for the parameters specified above. {Note that, even if the input for $F(r)$ and $G(r)$ are rotational invariant solutions of the effective theory, the envelope of the $8a_0$ charge density modulation is elongated in the $x$-direction. This is because in $H_{tr}$ the $C_4$ rotational symmetry is explicitly broken by the unidirectional PDW term.}
{The $4a_0$ stripe pattern is also discernible in this data, and it comes primarily from the regime closer to the vortex core than does the $8a_0$ peak.}

From $H_{tr}$ we can also compute the LDOS with a $d$-form factor, which probes the $xy$-anisotropy of  bond-centered density.\cite{Fujita-2014} For a CuO$_2$ plane in a cuprate HTSC, this can be directly measured via the charge density on the oxygen sites; in our single-band model, this corresponds to computing the LDOS using the spatial profile of
\begin{align}
\rho_d (\vec{r})= &\frac{1}{4}\left[\psi^*(\bm r)\psi(\vec{r}+\vec{a}) +\psi^*(\bm r)\psi(\vec{r}-\vec{a}) \right.\nonumber\\
&\left.-\psi^*(\bm r)\psi(\vec{r}+\vec{b})-\psi^*(\bm r)\psi(\vec{r}-\vec{b}) + c.c.\right],
\end{align}
where $\vec{a}$ and $\vec{b}$ are lattice vectors, and $\psi(\bm r)$ is the wavefunction of an eigenstate of energy $E$.

With the same parameters, the computed LDOS with a $d$-form factor is shown in Fig.\ \ref{fig:3}. We see that the $\bm Q$ and $2\bm Q$ peaks in the LDOS with $d$-form factor 
{ are somewhat weaker than those} with $s$-form factor. If these form factors had a symmetry  meaning (for example, {if they are distinguished by mirror} reflection in the diagonal direction), then we would have expected $s$-form factor for the LDOS only. However, diagonal reflection symmetry here is 
 broken by the charge stripes {and} by the vortex configuration. Therefore $s$-form factor and $d$-form factor components are always mixed.

The Fourier transform of the images in Figs.\ \ref{fig:2}, \ref{fig:3} are shown in Figs.\ \ref{fig:4} and \ref{fig:5}. We have only used the central $\frac{1}{2}\times\frac{1}{2}$ parts of the images, to exclude the rough edges. From the Fourier transforms, it is clear that the $8a_0$ stripe has dominantly an  $s$-wave form factor. While it is difficult to quantify due to finite size effects, from Fig.\ \ref{fig:4} we see that the $8a_0$ charge modulation has a sharper peak than the $2\Q$ one, i.e., has a larger correlation length, 
consistent with our analytical results.

 \begin{figure}[hbt]
 \includegraphics[width=\columnwidth]{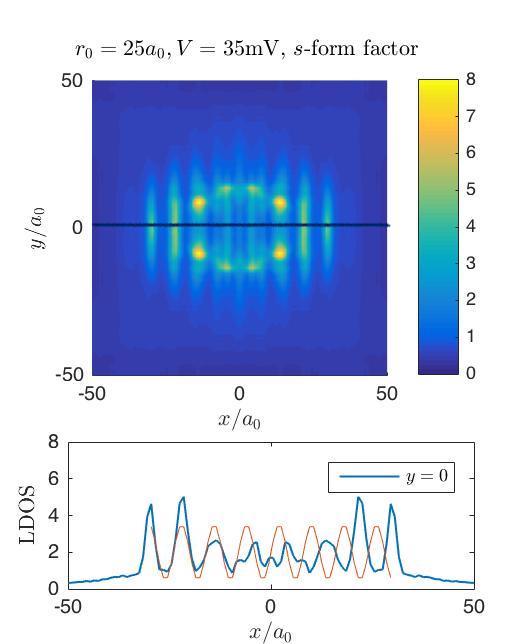}
 \caption{{ Charge density as a function of position in the neighborhood of an isolated vortex.  Notice the $\pi$-phase shift of the $8a_0$ stripes across the vortex center.  The lower panel shows the local density along the solid black line in the upper panel.
 Here the global phase of the SC order parameter is 
 chosen so that $\Delta(\bm r)$ is real 
 along the $x$ axis,  and the PDW order parameter is real. The red curve in the bottom panel is 
 a sine-function added to help visualize the $\pi$-phase shift.}
  }
  \label{fig:6}
 \end{figure}

\subsection{The structure of the $8a_0$ charge density modulation}
\label{subsec:pi}

Let us go back to $\tilde\rho_{\Q}$ defined in Eq.\ \eqref{eq:5}. 
For an \emph{isolated} vortex, we have considered the energetically favorable configuration where $\Delta$ has a phase winding around the vortex core, while $\Delta_{\Q}$, and the $8a_0$ stripe order $\tilde\rho_{\Q}$ emerges as the 
{ a composite} order of $\Delta$ and $\Delta_{\Q}$. A direct consequence of this is that 
across the center of the vortex, the 
$8a_0$
charge density modulation should exhibit a $\pi$-phase shift.
This feature is directly observed in the numerics, as we show in Fig.\ \ref{fig:6}. 

The argument for the $\pi$ phase shift of the $8a_0$ stripe only relies on the structure of the SC vortex, i.e., the fact the SC phase differs by $\pi$ across the vortex center. Thus, this feature should be robust for any configurations of the meron solution (whose PDW components differ by either a global phase or a translation, {\it i.e.}, by $\phi_{cdw}$ or $\phi_{pdw}$). We have indeed verified this, and present the results in Appendix \ref{app:cosQr}.

 It would very interesting to see whether the same feature can be observed in STM experiments near an isolated vortex too. It would provide 
  strong {(phase sensitive)} evidence for the PDW origin of the $8a_0$ stripe. On the other hand, if multiple vortices are close to each other, the phase winding around each vortex becomes distorted, then one would no longer expect this feature to be robust.

 \section{Broader implications for underdoped cuprates}
\label{sec:phasediagram}

\begin{figure}[hbt]
\includegraphics[width=\columnwidth]{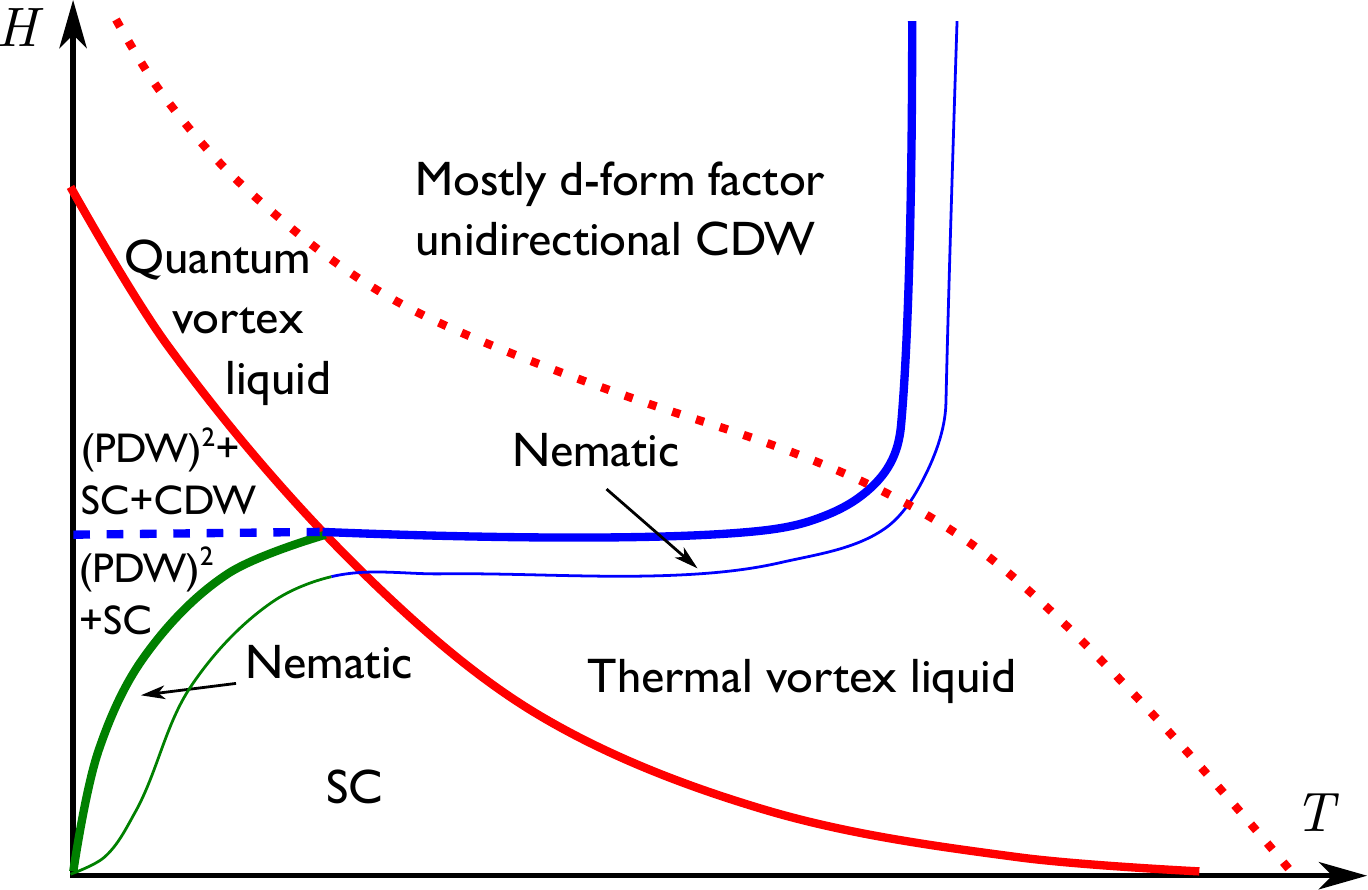}
\caption{A putative phase diagram for underdoped cuprates as a function of magnetic field $H$ and temperature $T$
in the absence of quenched disorder.
The solid red line marks a transition to a low-temperature/low-field $d$-wave SC vortex crystal phase, and the dotted red line represents a crossover region below which SC fluctuations are significant. Marked by the heavy green line, inside the vortex crystal phase, PDW fluctuations appear inside the SC vortex halo, which are not phase coherent across different vortices but induce a long-range CDW order with  primarily $s$-form factor. The 
explicit analysis 
of the current work applies to the low field region of this phase.
In the high-field/low-temperature region bounded by the heavy blue line, a long-range, unidirectional CDW order develops with a dominantly $d$-form factor. The two CDW phases at low fields and high fields are weakly coupled to each other via a commensurate lock-in. The thin green and blue lines denote a transition into a vestigial nematic phase, with lattice rotational symmetry broken but translational symmetry associated with the CDW order restored.}
\label{fig:phase}
\end{figure}

  As mentioned above,  enough of the  theoretically expected structure has been recently seen in STM studies of the vortex state of BSCCO-2212, that it is reasonable to conclude (as is argued in Ref.~\onlinecite{STM-8a}), that PDW order in the halos is an experimental reality.  In this section, we speculate on the implications of this for the understanding of the cuprates more broadly, in an attempt to place the present results in context.  Specifically, in Fig. \ref{fig:phase} we show a schematic phase diagram of a putative disorder-free hole-doped cuprate in the temperature-magnetic field plane, where to be concrete we have drawn the basic phase diagram with experimental results on underdoped YBCO-123 with doped hole concentration in the range $p \sim 0.1 - 0.13$ in mind.  The explicit calculations in the present paper apply directly only to the low field, low temperature regime, in the bottom portion of the phase denoted (PDW)$^2$+SC.

{\bf The SC phase:}  The high-temperature phase is not superconducting and is translationally invariant. 
The red line\cite{Ramshaw-2012} is the thermodynamic $H_{c2}$ line that marks the boundary of the $d$-wave SC  (vortex crystal) phase that exists at low enough fields and temperatures. 
(For graphical simplicity, we have ignored the existence of a Meissner state, i.e. we assume that $H_{c1} \approx 0$.)  

{\bf The SC + (PDW)$^2$ phase:}  Since at low fields, the vortex halos are non-overlapping, 
  the associated PDW order we have computed as a saddle-point of an effective action is presumably thermally disordered (unless pinned by impurities) down to low enough temperatures 
  that  the coupling between neighboring halos is large enough to produce long-range coherence.  The considerations governing the interactions between neighboring halos -- which are essentially identical to those discussed in Ref.~ \onlinecite{Kivelson-2002} -- lead to the heavy green phase boundary that marks the point at which the induced CDW correlations lock between different halos to macroscopic distances.  The resulting  phase has coexisting long-range CDW and SC order, although especially at low fields,  this order is likely to be quite easily destroyed by even rather minimal quenched disorder.  
  The magnetic field necessarily induces vortices in the superconducting phase associated with the PDW, but as a corollary of our results, these PDW-vortices lie preferentially in the region far from the PDW halos where the PDW order parameter is vanishingly small.  Thus,  the PDW vortices are likely  melted at all accessible temperatures.  It is therefore  only the induced CDW orders, $\tilde \rho_{\bf Q}$ and $\tilde \rho_{2\bf Q}$, that actually order in this phase.  We label the phase (PDW)$^2$+SC on the phase diagram to indicate that the coexisting orders are a usual, d-wave SC order and a CDW order that is associated with a  harmonic of the PDW order.  The CDW in this phase has a dominantly $s$-wave form factor.

{\bf The CDW phase:}  At high fields beyond the boundary of the SC phase, ultrasound experiments\cite{leboeuf-2012} in  {\YBCO} have detected an additional thermodynamic phase transition marked by the heavy blue line in the figure;  on the basis of NMR\cite{Wu-2011,Wu-2013} and high field X-ray diffraction experiments,\cite{Jang-2016,chang-2012,gerber} this transition has been identified with the onset of long-range unidirectional CDW order.  The existence of small electron pockets detected in quantum oscillation\cite{doiron-leyraud-2007,LeBoeuf-2008,Sebastian-2008,Sebastian-2015,Ramshaw-2015} experiments are widely accepted as reflecting the Fermi surface reconstruction produced by the high-field CDW order.  Short-range correlated versions of the same order -- presumably pinned by quenched disorder -- have been observed in a larger range of temperatures and (low) magnetic fields by X-rays\cite{chang-2012,Ghiringhelli-2012,blackburn-2013,Blanco-Canosa-2014,Comin-2015,Achkar-2016,Jang-2016}} and NMR\cite{Wu-2011,Wu-2013,julien-2015}.  The existence of closely related CDW correlations have been inferred in a variety of STM (and M-EELS) measurements on {\BSCCO}, as well.\cite{lawler-2010,silva_neto-2012,Silva_Neto-2014,Hamidian-2015,Vig-2015,Mesaros-2016}  The identification of   the low field CDW correlations as disorder pinned versions of  the high field CDW order is strengthened by the observations that they have the same in-plane ordering vector, that both forms of order appear to favor unidirectional (stripe) over bidirectional  (checkerboard) ordering.  Moreover, there is some evidence that  they probably both  have a dominantly d-wave form factor.

{\bf One CDW or Two:}  A key issue of perspective is whether the high field CDW  is essentially a separate order, $\rho_{2{\bf Q}}$, or a parasitic consequence of a much stronger PDW ordering tendency.  
In drawing our phase diagram, we have assumed that the PDW is always (slightly) subdominant to the SC order, and so have identified the high field phase as being primarily a CDW. In contrast, the alternate view -- that PDW order is much more stable at high magnetic fields than the SC -- was adopted in Ref.~\onlinecite{agterberg} and still more emphatically in Ref.~\onlinecite{lee-2014}.  The fact that the  $\tilde \rho_{2{\bf Q}}$ order in the vortex halos has the same ordering vector as the disorder pinned CDW correlations seen in many older STM studies  suggests that they are related.  However, this can be accounted for by a mutual commensurate lock-in between the two forms of CDW order, as in Eq.~\eqref{lockin}.  Conversely, as stressed above, the low field CDW has a predominantly $d$-wave form factor while the field induced order in the halos is dominantly $s$-wave.  {Moreover, the field induced order  is also almost perfectly particle-hole symmetric in its modulation intensity, 
while the zero-field CDW which is 
largely antisymmetric.} Moreover, one would generally expect that in a globally superconducting state, the PDW-induced CDW order should exhibit ordering vectors at both ${\bm Q}$ and $2{\bm Q}$;  indeed, in any state that has dominant SC order and subdominant PDW order, the  ${\bm Q}$ peak should be stronger than the $2{\bm Q}$.  No clear evidence of a ${\bm Q}$ peak has been reported in the low field STM on {\BSCCO}.  {All} these observations clearly favor the idea that there are two distinct forms of CDW order that are (weakly) coupled so as to be mutually commensurate where they coexist.

There is further evidence against interpreting the high field CDW phase as originating from PDW order.  In the first place, over most of the region of $T$ and $H$ in which CDW order is shown, the system is not superconducting, and (beyond a crossover line - the dashed red line in the figure) does not show any clear evidence of identifiable short-range superconducting coherence.  The fact that the CDW transition temperature, $T_{\rm CDW}$, (the heavy blue line) is essentially field independent at high fields -- which follows directly from the ultrasound and NMR experiments in {\YBCO} -- further corroborates this.  It is generally accepted that the magnetic field does not have much direct effect on CDW ordering;  rather, increasing field suppresses local superconducting order, with the result that to the extent that CDW and SC orders compete, a high field will indirectly enhance the CDW order.  A corollary of this is that once the SC is sufficiently weak that it no longer competes with CDW order, the field ceases to affect the CDW order.  Conversely, if the CDW order was induced by PDW order, it is difficult to see why $T_{\rm CDW}$ would not be a strongly decreasing function of increasing $H$.  Finally, the presence of quantum oscillations with seemingly standard Onsager periodicity as a function of $1/H$ and a temperature dependence that is very accurately given\cite{sebastian-2010} by the Lifshitz-Konsevich form, is difficult to reconcile with the existence of a PDW, even though a PDW does support Fermi pockets of an appropriate size.\cite{berg-2008a,berg-2009,baruch-2008,zelli-2011,zelli-2012,lee-2014,Vafek-2013}

{\bf Vestigial nematic order:}  As we have assumed that both the preferred CDW and the PDW order are unidirectional, there is the possibility that fluctuations will cause a two-step melting of the density wave order.  
Upon increasing temperature, the lattice translational symmetry that is broken by the density-wave orders is restored by thermal fluctuations. However, the lattice rotational symmetry breaking, being a $\mathbb{Z}_2$ Ising degree of freedom, is more robust against thermal fluctuations and survives to  higher temperatures~\cite{schmalian, nie-2013, Wang-2014}. The intermediate phase with broken lattice rotational symmetry  but intact translational symmetry is an Ising
nematic bounded by a nematic critical temperature shown as the thin blue and green lines in the figure.  While in the absence of disorder, such nematic phases tend to occur in very narrow slivers of the phase diagram, the presence of quenched disorder can greatly enhance their importance.\cite{nie-2013}  Evidence of the existence of such nematic phases has been reported in various experiments, including STM experiments\cite{kivelson-2003,lawler-2010} in {\BSCCO} and neutron scattering\cite{hinkov-2007} and Nernst effect   \cite{Daou2010} measurements in {YBCO}.

{\bf More than one vortex crystal phase:}  To the extent that the two forms of CDW are approximately distinct, a crossover must occur within the coexistence phase, as shown by the dashed blue line in the figure. Near the boundary of the SC phase, the CDW that coexists with SC must look increasingly like the high-field CDW, while well below the dashed line, its origin as a consequence of a primary tendency to PDW order must become increasingly pronounced.  Under some circumstances this dashed line could correspond to a sharp phase transition;  for instance, if the PDW were bidirectional and the CDW were unidirectional, then the dashed line would likely correspond to a first order transition line, at which a coexisting nematic order parameter first occurs.

{\bf Superconducting fluctuations:}
The red dotted line represents a (not precisely defined) crossover to a vortex liquid. Under the assumption that the primary way magnetic fields couple to CDW order is indirectly, through its effect on the SC order, the portion of this curve that lies within the CDW ordered phase can be inferred indirectly from the field-dependence of the CDW order:
Below the dashed curve, the strength of the CDW order is an increasing function of $H$, while above the dashed curve (where by some measure SC order has been sufficiently suppressed) the strength of the CDW is essentially field independent.  While there is ample evidence that SC fluctuations survive at low fields to   temperatures much farther above the zero field $T_c$ than in any conventional superconductor, there is considerable controversy about exactly how broad a range, and on how to define the appropriate crossover scale.  We do not attempt to resolve this issue here.  However, this is likely a place where the existence of a slightly subdominant PDW order parameter plays a significant role in a broader range of the phase diagram than that in which actual PDW order occurs.  Precisely because PDW order arises in the vortex core, the vortex core energy is much less than it would otherwise be, leading to precisely the ``large" and ``cheap" vortices that are required to produce a large fluctuational regime.

{\bf Comparison with the theory in Ref.~\onlinecite{Agterberg-Garaud-2015}:}  Compared with the phase diagram proposed in Ref.~\onlinecite{Agterberg-Garaud-2015}, our phase diagram shown in Fig.\ \ref{fig:phase} differs in several important aspects. First, in our phase diagram, the PDW phase is \emph{always} subdominant to the uniform $d$-wave SC, and it only appears in the vortex phases below $H_{c2}$. In Refs.~\onlinecite{lee-2014,Agterberg-Garaud-2015} it has been proposed that PDW order  survives to a higher magnetic field than uniform SC does. However, the microscopic justification of this scenario is unclear, particularly given that at zero field uniform SC is clearly the dominant order. Second, in Refs.~\onlinecite{Agterberg-Garaud-2015}  it is assumed that the PDW order is bi-directional.
 This is not a major difference from a theoretical perspective -- the difference between the two cases is encoded in the sign of $\gamma$ in Eq. \eqref{dotdotdot}. However,  the experimental data from STM and X-ray suggests that the zero field CDW order is unidirectional. While the interpretation of the patterns in the vortex cores is still unsettled,  this observation certainly is suggestive of a tendency to  uni-directional PDW order.
(Notice,  that disorder always blurs the distinction between unidirectional and bidirectional charge order \cite{robertson-2006}).
Third, as already discussed, our proposed phase diagram assumes that there are two microscopically distinct forms of CDW ordering --- one with primarily a $d$-wave form factor and the other with primarily a $s$-wave form factor.  

{\bf The case of the dominant PDW:} 
Given that PDW and SC are found to be very close in energy in numerical studies of $t-J$, \cite{Himeda-2002,Raczkowski-2007,corboz-2014}
it is also plausible to consider a phase diagram in which PDW order is slightly dominant to the uniform SC order. Indeed, in {\LBCO} near $1/8$ doping and in underdoped {\LSCO} in magnetic fields, a dominant PDW order can naturally explain the dynamical layer decoupling in transport properties.\cite{Li-2007,Tranquada-2008,fradkin-2014} In this case it would be interesting to consider the vortex states with uniform SC component in the halo of half vortices or full vortices of PDW order.  More generally, in a magnetic field, evidence of an induced PDW superconductor has been observed in various members of the LCO (214) family of cuprates. 
This case is discussed in some detail in Ref. \onlinecite{fradkin-2014}. 
 
 {\bf Still more exotic phases:}  Going back to the lower edge of the phase diagram, we have argued that when the vortices are far separated, the superconducting phase associated with the PDW will generically be disordered.  However, as the vortex halos approach each other, one could imagine a phase in which long-range PDW order coexists with long-range SC order.  Here, there would be a two-component vortex lattice -- one set of vortices associated with the usual d-wave SC and the other with the PDW superconductor.  Indeed, if the PDW were bidirectional, still more complex phases could be imagined.  These phases would have a rich variety of novel Goldstone modes and topological excitations.  (Some of this has been discussed in Ref. \onlinecite{agterberg}.)  Then, of course, with all these order parameters, the possibility of phases with various forms of vestigial or composite order is combinatorially large.

 \section{Summary}
 \label{sec:discussion}
 
 In this paper we have developed an effective model for SC and PDW orders, 
 in which the intertwining of these order parameters gives rise to different types of charge density modulations. We solved the effective theory in the vicinity of a SC vortex, and found that the PDW order is locally enhanced in a  vortex halo. The induced charge modulations have both period $\sim 4a_0$ and $\sim 8a_0$ components. While the  $\sim 4a_0$ charge order has been ubiquitously discovered in cuprate HTSC's,  direct detection of the $\sim 8a_0$ near SC vortex halos  provides strong evidence for the elusive PDW order. 
  
The following additional features of the solution are largely independent of microscopic details.  
 The $\sim 8a_0$ charge modulation near a vortex core has a larger spatial profile than that of the $\sim 4a_0$ one. 
The $\sim 8a_0$ charge modulation  has a 
predominantly $s$-wave form factor.
For an isolated SC vortex, 
the $\sim 8a_0$ charge stripe  
exhibits a $\pi$-phase shift  across the vortex center. 

We connected our results on vortex PDW states to a putative phase diagram for an underdoped cuprate in an external magnetic field. In the region where PDW coexists with SC in the vortices,
in addition to the ongoing STM measurements already mentioned,\cite{STM-8a} it would be extremely illuminating to look for a related subharmonic CDW peak in X-ray diffraction at intermediate fields less than but comparable to $H_{c2}$, where the STM results suggest that uniform SC and PDW orders coexist in an inhomogeneous fashion.

{\em Note added}: {After this work was completed, we received an advanced draft of a paper by M. R. Norman that discusses the possible connection between the PDW and quantum oscillation experiments.\cite{norman-2018} After completion of this work we have also received a draft paper~\cite{palee-new} from Patrick A. Lee that analyzes a similar problem. 
We thank the authors for sharing their unpublished work with us.}

\begin{acknowledgments}
{We thank D.\ F.\ Agterberg, A.\ V.\ Chubukov, M.\ H.\ Julien, P.\ A.\ Lee, B.\ Ramshaw, S.\ Sachdev, S.\ Sebastian, T. Senthil and J.\ Tranquada for inspiring discussions.}
This work was supported in part by the Gordon and Betty Moore Foundations EPiQS Initiative through Grant No. GBMF4305 (YW), and Grant No GBMF4544 (J.C.S.D and M.H.H), by the National Science Foundation grant DMR-1725401 at the University of Illinois (EF), by  EPSRC under Grant EP/G03673X/1 (SDE), and by  the Department of Energy, Office of Basic Energy Sciences, under contract No. DE-AC02- 76SF00515 at Stanford (SAK), and contract No. DEAC02-98CH10886 at Brookhaven National Lab (J.C.S.D), and the Karel Urbanek postdoctoral fellowship at Stanford University (SDE).
\end{acknowledgments}

 \begin{figure}[hbt]
 \includegraphics[width=\columnwidth]{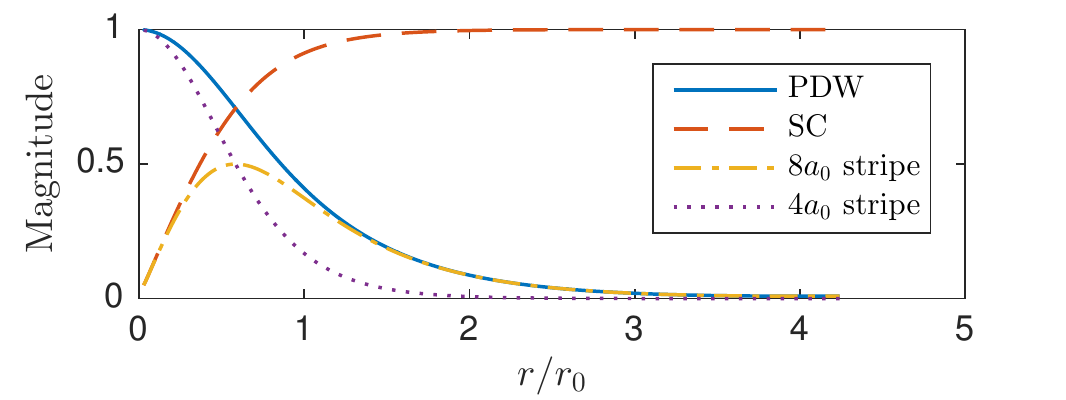}
 \caption{{Numerical solution to the meron-like configuration of the order parameters near a vortex halo in the case in which the phases of PDW and SC order parameters are locked and \emph{both} wind by $2\pi$.  We have used $\bar \kappa = 0.5 \kappa$, and $\lambda=0.5\rho$.  The magnitudes of $\tilde\rho_{\bm Q}$ and $\tilde\rho_{2\bm Q}$ are computed from Eqs. \eqref{eq:5} and \eqref{eq:6} with coefficients $B_{1}=B_2=1$.}
}
  \label{fig:1a}
 \end{figure}

\begin{appendix}
 \section{Phase-locked solution in a vortex halo}
 \label{app:lock}

 In this Appendix we search for solutions of the saddle point equation with phase locking between $\Delta$ and $\Delta_{\pm Q}$, {{\it i.e.} when there are coincident vortices in both the uniform SC and the PDW components of the order.}
{We restrict our attention to the case in which $\rho > \lambda$;  this is a necessary condition to insure that the global, vortex free minimum of the effective free energy corresponds to a uniform SC state with no PDW component.}
 
To proceed, we still use the parametrization $|\Delta|=\sin \alpha$ and $|\Delta_{\pm Q}| = \frac{1}{\sqrt{2}}\cos \alpha$, ($0<\alpha<\pi/2$), while their phases are fixed by the boundary condition and phase locking condition. In this configuration the effective free energy becomes
 \begin{align}
 \label{eq22}
&S_{\rm I}[\alpha] \nonumber\\
=& \pi\kappa \int r dr \lf[ \lf(\frac{d\alpha}{dr}\rt)^2 + \frac{1}{r^2} -\frac{\lambda}{\kappa} \sin^2\alpha\cos^2\alpha +\frac{\rho}{\kappa} \cos^2\alpha \rt],
\end{align}
where the $1/r^2$ term comes from the $2\pi$ phase winding of both the PDW and SC components.
This free energy is clearly minimized by an $r$-independent solution with
\begin{align}
\alpha=\frac{\pi}{2},
\label{eq:gs}
\end{align}
i.e., the solution is a bare SC vortex with no PDW component. From this result, the meron solution is clearly more energetically favorable.

 \begin{figure}[hbt]
 \includegraphics[width=\columnwidth]{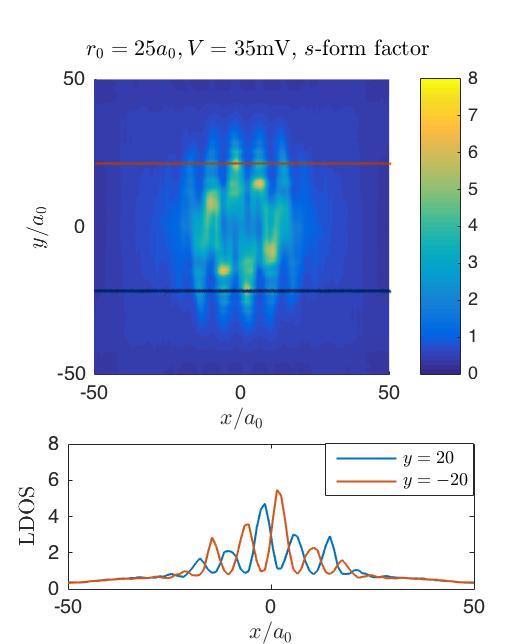}
 \caption{Same as in Fig. \ref{fig:6}, but with the global phase choice $\phi_{cdw}=\pi/2$ and $\phi_{pdw}=\pi/2$.  The lower panel shows the local density along the two cuts shown in the upper panel.}
  \label{fig:7}
 \end{figure}

However, after a closer look, the complete absence of PDW component in our solution is an artifact of setting $
\delta \kappa=0$ and $\tilde \kappa=0$ in Eq.\ \eqref{dotdotdot}. Indeed, if the superfluid density for PDW order is lower than that for SC, (which is likely to be true in realistic cases), then it is energetically more favorable to  
have a PDW component 
in 
a region near the vortex core.  In this case, we again expect a meron-like solution, 
 only now the  PDW phase is locked to the SC phase and also winds.
 
To verify this, we have numerically minimized the corresponding effective free energy that includes $
\delta \kappa$,  
 \begin{align}
 \label{eq23}
 S_{\rm I}[\alpha]=& \pi\kappa \int r dr \lf[ \lf(\frac{d\alpha}{dr}\rt)^2 + \frac{1+(
 \delta\kappa/\kappa) \sin^2\alpha}{r^2}\right.\nonumber\\
&\left.~~~~~~~~~~~~-\frac{\lambda}{\kappa} \sin^2\alpha\cos^2\alpha +\frac{\rho}{\kappa} \cos^2\alpha \rt],
\end{align}
again by mapping it to a classical mechanics problem, and have indeed found such a solution. The size of the vortex halo $\tilde r_0$ is a given by
 \be
 \tilde r_0 = r_0 f\(\frac{\lambda}{\rho}, \frac{ \delta \kappa}{\kappa}\),
 \ee
where $f(x,y)$ is a non-universal function.  For specific values $
\delta \kappa = 0.5 \kappa$, and $\lambda=0.5\rho$, we show a representative plot of the solution and their induced charge density modulations in Fig.\ \ref{fig:1a}. The magnitudes of PDW and SC components behave qualitatively the same as those shown in Fig.\ \ref{fig:1}, however their phases are locked and both wind by $2\pi$.

Since the PDW configurations with or without a phase winding have distinct topology, they cannot smoothly deform to each other. To determine the optimal configuration, one just needs to compare their respective energies. For the phase-locked solution, the vortex halo reduces the vortex core energy by, to leading order in the ultraviolet cutoff,
\be
\Delta E' = -\pi 
\delta \kappa \ln ({\tilde r_0}/{a}).
\ee

On the other hand, including a nonzero $\delta \kappa$ does not alter the behavior of the meron solution with a uniform PDW phase much; to leading order the only change is Eq.\ \eqref{eq19} becomes $\Delta E = -\pi (\kappa+
\delta\kappa) \ln (r_0/a)$. Since $\kappa>0$, comparing the energies of the vortex halo, it remains true that the meron solution with a uniform PDW phase is optimal.\\

 \begin{figure}
 \includegraphics[width=\columnwidth]{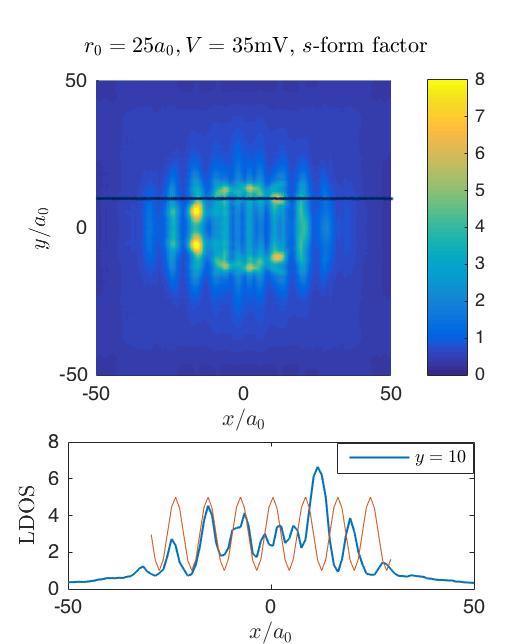}
 \caption{
 Same as in Fig. \ref{fig:6}, but with the global phase choice $\phi_{cdw}=\pi/2$ and $\phi_{pdw}=\pi/2$.  The lower panel shows the local density along the two cuts shown in the upper panel.}
  \label{fig:8}
 \end{figure}

\section{Local density of states for different choices of $\phi_{cdw}$ and $\phi_{pdw}$}
\label{app:cosQr}

In Fig.\ \ref{fig:7}, we show the LDOS for $\phi_{cdw}= \pi/2$, and $\phi_{pdw}=\pi/2$, while keeping all other parameters the same as those in Sec.\ \ref{sec:ldos}. In the lower panel Fig.\ \ref{fig:7}, we show the local density long two cuts at $y=\pm 20 a_0$. They indeed differ by a phase difference $\pi$, just as we argued in Sec.\ \ref{subsec:pi}.

The same data is shown in Fig.\ \ref{fig:8}, only for $\phi_{cdw}= 0$, and $\phi_{pdw}=0$. We show the phase difference of $\pi$ across the vortex center in the lower panel.

\end{appendix}

\bibliographystyle{apsrev4-1}
\bibliography{references.bib}

 \end{document}